\documentclass[useAMS,usenatbib]{report}
\usepackage{amsmath,amssymb,wasysym}
\usepackage{booktabs}
\usepackage{authblk}
\usepackage{delarray}
\usepackage{graphicx}
\usepackage{tabularx}
\usepackage{hyperref}

\newcommand{\bv}{\begin{verbatim}}
\newcommand{\ev}{{ \end{verbatim} }}

\newcommand{\nicefrac}[2]{\leavevmode\kern.1em
            \raise.5ex\hbox{\the\scriptfont0 #1}\kern-.1em
      /\kern-.15em\lower.25ex\hbox{\the\scriptfont0 #2}}

\begin{document}


\title{{\huge Fitting galaxy spectra with STECKMAP:
a user guide}}

\author{P. Ocvirk}
\affil{Observatoire Astronomique de Strasbourg, 11 rue de l'Universit\'e, 67000 Strasbourg, France. \\
\vspace{0.2cm}
pierre.ocvirk@astro.unistra.fr}





\maketitle












\chapter*{What is STECKMAP?}

STECKMAP stands for STEllar Content and Kinematics via Maximum A Posteriori likelihood. It is a tool for interpreting galaxy spectra in terms of their stellar populations, through the derivation of their star formation history, age-metallicity relation, kinematics and extinction. From these, a number of integrated quantities are also computed, such as luminosity-weighted age and metallicity, and mass-weighted age and metallicity (respectively, LWAge, LWZ, MWAge, MWZ).

To do so, the observed spectrum is projected onto a temporal sequence of models of single stellar populations, so as to determine a linear combination of these models, that fit the observed spectrum best (via a penalized $\chi^2$). The weights of the various components of this linear combination indicate the stellar content of the population. This procedure is regularized using various penalizing functions.
 The principles of the method are detailed in \cite{stecmap,steckmap}.

The STECKMAP software package is public and freely available at\\ \url{https://github.com/pocvirk/STECKMAP}. A number of authors have already adopted it and use it in their daily research. Examples of a variety of applications of STECKMAP can be found in \cite{pat2007,koleva08,sharina09,koleva09,ocvirk2010,pat2011}

This user guide aims at accompanying the user through the setup and first runs of STECKMAP. The last chapter will help the user to understand and improve his results and experience with the code.

\tableofcontents

\chapter{STECKMAP setup}
One needs to install Yorick and STECKMAP separately.

\section{Installing yorick}
You will thus need to first install Yorick.
Under ubuntu or mac instructions can be found online.
This should in principle be painless, especially using the pre-compiled binary distributions. Yorick is very light (much lighter than the
STECKMAP package itself) so you wont have to worry about disk space. I recommend installing Yorick in a directory such as
{\tt \$HOME/Yorick/}

Successful installation should result in the yorick executable to have a full path similar to:

\$HOME/Yorick/yorick-2.1/yorick/yorick
(for yorick-2.1 that is).
This note provides few guidelines for installing yorick. Refer to yorick's README for more, as it is more complete. On Macs, yorick can be installed through fink.
I recommend using yorick through emacs. This allows for 2 very cool things to happen:
\begin{itemize}
\item{enables emacs code editing features: colors, indentation, etc... i.e. a helpful coding environment.}
\item{keyboard shortcuts: the buffer you are currently editing can be executed with Ctrl-X Ctrl-S.}
\end{itemize}

Follow the yorick.el instructions to set this up. You will need to make a straightforward modification of your ~/.emacs. and inside yorick.el dont forget to provide the full path to the yorick executable (search for the string "yorick-executable-name" and replace the adjacent string "yorick" with the full path to the executable).
You can check if yorick is correctly installed by launching it typing 
\begin{verbatim}
bash-3.2$ yorick
 Copyright (c) 2005.  The Regents of the University of California.
 All rights reserved.  Yorick 2.1.05 ready.  For help type 'help'
> include, "demo2.i"
> demo2
\end{verbatim}
A cool animation should ensue, greeting you to a successful install of Yorick.

\section{Installing STECKMAP}
You can find the latest package from here:\\
\url{https://github.com/pocvirk/STECKMAP/}
Download the latest version or just follow git instructions to clone the repository.
Make sure to clone it into your Yorick directory (\$HOME/Yorick/).

In order to be able to relocate the code easily, an environment variable is used to locate the directory where STECKMAP is installed. Its name is {\tt STECKMAPROOTDIR}.
I usually install Yorick and STECKMAP in {\tt \$HOME/Yorick/}, so that I can set (for a bash shell):
\begin{verbatim}
export STECKMAPROOTDIR=$HOME/Yorick/
\end{verbatim}

In this setup, this rootdir looks like this:
\begin{verbatim}
bash-3.2$ cd $STECKMAPROOTDIR
bash-3.2$ ls
STECKMAP
yorick-2.1
\end{verbatim}

\section{Checking the successful installation with an example}
Launch yorick by typing "yorick" on the shell command line.
Once yorick is launched, load STECKMAP by typing 
\begin{verbatim}
> include, "STECKMAP/Pierre/POP/sfit.i"
\end{verbatim}

You will find a couple of example data in Yorick/Pierre/POP/EXAMPLES/.
They are provided in order to give the user a taste of what can be done
and how to proceed. Follow this small tutorial.

STECKMAP is a tool package aiming at constraining the stellar content and
kinematics from galaxy spectra. To do so, you need two things:
\begin{itemize}
\item{DATA}
\item{Models against which to compare your data}
\end{itemize}

The detailed use of the function will be explained in the further chapters. For the moment we just want to check STECKMAP is installed properly.

\subsection*{Data}

See the path and name of an example file I set up by
typing
\begin{verbatim}
> fV
\end{verbatim}

fV is just a variable storing the path and name of the example file. To
make things easier, it is loaded by default.
Convert the example file by typing:
\begin{verbatim}
> a=convert_all(fV)
\end{verbatim}

this will create the .pdb file, plot the data, and write the redshdift of
the example galaxy if it is supplied in the fits keywords. Otherwise it
will just assume 0, as in this case.
Note that it will plot the 2d plate and the spectrum obtained by summing
the whole 2d spectrum in the spatial direction. This is the default
behaviour of convert\_all.

A model basis is a structure as defined in "Pierre/POP/base\_struct.i". This structure contains the sequence of SSPs in time and metallicity. As such, it is a data cube. The structure contains this cube and also the wavelengths, the resolution, the metallicity scale, and the ages of wach element of the basis.

\subsection*{SSP basis}

To generate a basis, use the function bRbasis3:
\begin{verbatim}
> b=bRbasis3()
\end{verbatim}

This will call bRbasis3 with all the defaults arguments. To get help and see information about the various arguments and their default values you can type
\begin{verbatim}
> help, bRbasis3
\end{verbatim}
or
\begin{verbatim}
> info, bRbasis3
\end{verbatim}

Here the default is a PEGASE-HR sequence of SSPs  with 10 age bins from 10Myr to 20 Gyr with Salpeter IMF and Padova tracks. It is flux-normalized by default (see the STECMAP paper for details). PEGASE-HR has a resolution R=10000 and the wavelength coverage is $4000-6900$ {\AA}. All these values can of course be changed when calling bRbasis3.

To have an idea of the looks of the  basis you just generated, type
\begin{verbatim}
> ws  
\end{verbatim}
this is to clear the display
\begin{verbatim}
> plb, b.flux(,,1),b.wave
\end{verbatim}
This will result in a nice plot of the basis for the first metallicity of the basis. It can be printed by typing:
\begin{verbatim}
> b.met(1) 
\end{verbatim}
since the metallicities are stored in b.met. Note that for computational reasons, the metallicities are renormalized. Hence, you can read the original metallicity of the first constant-metallicity slice of the SSP cube b.flux(,,1) by typing:
\begin{verbatim}
> Zrescalem1(b.met(1))
\end{verbatim}
it should be 0.05.

\subsection*{Fitting engine}

The fitting engine is called with the sfit command, which takes as arguments, the data, the basis and a bunch of options. Type 
\begin{verbatim}
> help,sfit
\end{verbatim}
to see the current documentation for sfit.
To fit the data, the structure of which is stored in the variable a, using the SSP basis, the structure of which is stored in the variable b, type:
\begin{verbatim}
> x=sfit(a,b,kin=1,epar=3,noskip=1,sav=1)
\end{verbatim}
The various options will be explained in the next chapter.

\chapter{Fitting your data}
\section{Preparing your data: convert\_all}
\label{s:convert}
Spectra of galaxies come in a variety of forms, 1D (such as SDSS), 2D (typical long-slit spectroscopy), 3D (data cubes), some of them have attached errors and masks (e.g. SDSS), others not.
STECKMAP on the other hand will fit only one spectrum at a time. 
Therefore the first step in the analysis is to convert the available data into a 1D spectrum in a format that STECKMAP understands. This can be done in the majority of cases with the {\tt convert\_all } routine.
Since even in the case of 3D spectroscopy spectra are collected on a 2D detector and even stored in 2D structures (wavelength, fiber number), {\tt convert\_all} can deal at the moment only with 1D and 2D provided spectra. The inputs and options can be obtained as 
\begin{verbatim}
> info,convert_all
func convert_all(filelist,cut=,noplot=,log=,z0=,SNR0=,wav=,
wavaxis=,xs=,xe=,hdu=,fsigm=,errorfile=)
\end{verbatim}

A typical call looks as follows:
\begin{verbatim}
> a=convert_all(fV,xs=100,xe=300,z0=500./300000.);
 NAXIS2>NAXIS1, assuming AXIS2 wavelength axis
 no redshift in fits header... will assume 0 or take user input
 no S/N in fits header... will assume 100 or take user input
 redshift 0.00166667
> 
\end{verbatim}

\subsection{Input}
The only mandatory input is the filename of the data file to be ingested {\tt filelist}. This must be a fits file following 
Note that it can be a list but it is usually more practical to convert and analyse spectra one by one.

\subsection{Options}

\begin{tabular}{lp{9cm}}
{{\tt {hdu=}}} & if fits file contains multiple header data units, specify number of hdu to read. \\
{\tt cut=} & filelist is cut at cut-th file, default is 10\\
{\tt log=} & log=1 enforces log wave sampling in case fits header information is inaccurate. \\
{\tt noplot=} & disables the plotting of the spectrum read (useful when remotely running a batch of spectra on a machine without an active X11 window with nohup for instance. default is noplot=0, so plotting happens.\\
{\tt  z0=} & if redshift is not provided in fits header, can be given by user as z0= \\
{\tt SNR0=} & same as z0 for global signal to noise ratio. Note that ideally it is better to supply a noise spectrum via {\tt errorfile=}.\\
{\tt errorfile=} & specify a name for an error file in order to fill the sigm vector. This will only work for 1D spectra right now.
\\
{\tt wavaxis=} & possible values are 1 or 2. Useful if data is provided as 2d frame, typical for long-slit spectroscopy. If no value is provided (default) we take as the wavelength axis as the axis with largest dimension. \\
{\tt xs=, xe=} & start and end of the stacking in the spatial direction, useful for long slit spectroscopy. \\
{\tt sigm=} & sigm=1 forces sigma=1 instead of d/(S/N).\\
\end{tabular}

\subsection{Output}
The direct result of the call is a structure containing various informations about the spectrum and the galaxy, and the various filenames of the results files {\tt .res}, the original data file {\tt .fits}, the intermediate data file {\tt .pdb}. This latter file is written by the routine and will be accessed by the fitting routine {\tt sfit}.
You can see the content of the result structure by typing:
\begin{verbatim}
> a
[galStruct(name="NGC4621",filename=
"/Users/Pedro/Yorick/STECKMAP/Pierre/POP/EXAMPLES/NGC4621-1.pdb",result_dir=
"Users/Pedro/Yorick/STECKMAP/Pierre/POP/EXAMPLES/",resfile=
"/Users/Pedro/Yorick/STECKMAP/Pierre/POP/EXAMPLES/NGC4621-1.res",wavesampling=
"LIN",redshift=0.00166667,SNR=100)]
> 
\end{verbatim}

\section{Preparing a sequence of SSP models: \\ bRbasis3}
\label{s:bRbasis}
There are several versions of the routine for generating the SSP basis. A basis is a datacube, the 3 dimensions being ($\lambda$, age, Z). Note that here in all of this user guide the term ``basis'' is used somehow abusively: because of the correlations between the model spectra at different ages and metallicities, such a family of vectors is usually {\em not} linearly independent and therefore it is not a basis in the linear algebra sense.
With this warning, we will stick to this nomenclature for practical reasons, and because it captures well the idea of ``projecting'' the observed spectrum on a set of model spectra. Note that the decomposition obtained in this way is unique only if the set of model spectra is linearly independent, i.e. is a basis. The most advanced at the moment is {\tt bRbasis3}. Information can be obtained by typing:
\begin{verbatim}
> info,bRbasis3
func bRbasis3(ages,&FWHM,&bab,mets=,nbins=,R=,wavel=,N=,basisfile=,dlambda=,
zr=,br=,inte=,navg=,list=,intages=,dirsfr=,filters=,s=)
> 
\end{verbatim}

A typical call looks as follows:

\begin{verbatim}
> b=bRbasis3([1.e8,2.e10],wavel=[3500.,7000.],basisfile="MILES",nbins=30);
\end{verbatim}

\subsection{Inputs}
The most important input here is the age ranges where the models will be computed. It is given as a 2-vector $[age_{min},age_{max}]$. The unit is years.

\subsection{Options}
\begin{tabular}{lp{9cm}}
{\tt nbins=} & Number of ages bins of the basis. \\
{\tt wavel=} & Wavelength range of the models (by default the broadest available range is taken. \\
{\tt R=} & The models can be broadened to an arbitrary spectral resolution to account for instrumental spectral resolution or Doppler broadening of the galaxy spectrum. Note that this depends on the assumed resolution for each model (can be checked by typing {\tt > b.R}. \\
{\tt basisfile=} & Reference of the models to be used. See table \ref{t:sspmodels} for a list of the models available with the code. \\
\end{tabular}

\begin{table}
\begin{tabular}{lrrr}
Code & Reference & age range (yr)& $[$Z/H$]$ \\ 
\hline
{\tt ``BC03''} & Bruzual \& Charlot 2003 & $10^{5}-1.7 10^{10}$ & [0.3,-2]\\
{\tt ``MILES''} & Vazdekis 2010 & $2.10^{7}-1.7 10^{10}$ & [0.2,-1.3]\\
{\tt ``PHR''} & Leborgne et al. 2004 &  $2.10^{7}-1.7 10^{10}$ & [0.2,-2]\\
{\tt ``GD05''} & Gonzalez-Delgado et al. 2005 & $2.10^{7}-1.7 10^{10}$ & [0.2,-2]\\
\end{tabular}
\caption{References of the SSP models available with STECKMAP.}
\label{t:sspmodels}
\end{table}

\section{Running the fitting engine: sfit}

This section details the call to {\tt sfit} and some of the options available for the fit. This is where most of the work is done, and therefore where most of the computation time will be spent.

\subsection{Inputs}
 The first mandatory argument of {\tt sfit} is the data structure containing the spectrum to be fitted and its metadata. This is the result of the command {\tt spdata=convert\_all(``myspectrum.fits'')} as explained in Sec. \ref{s:convert}. 
The second mandatory argument of {\tt sfit} is the spectral basis of models {\tt b}, as obtained from running {\tt bRbasis}, explained in Sec. \ref{s:bRbasis}
The simplest STECKMAP fit can then be obtained by issuing:
\begin{verbatim}
> x=sfit(spdata,b)
 ITER    EVAL     CPU [s]            FUNC             GNORM      STEPLEN
------  ------  ----------  -----------------------  ---------  ---------
     0       1    1.19e-02   1.9995698763700508e+04    8.3e+02    0.0e+00
   100     123    6.39e-01   7.3005684797028195e+01    5.9e+01    1.0e+00
   150     543    2.71e+00   7.2729881396562234e+01    1.6e+01    5.7e-18
warning: too many function evaluations (543)
 ITER    EVAL     CPU [s]            FUNC             GNORM      STEPLEN
------  ------  ----------  -----------------------  ---------  ---------
     0       1    4.78e-03   5.3944698728148160e+02    4.3e+05    0.0e+00
    82     462    2.32e+00   7.2666269919203131e+01    1.6e+01    8.5e-21
 ITER    EVAL     CPU [s]            FUNC             GNORM      STEPLEN
------  ------  ----------  -----------------------  ---------  ---------
     0       1    5.36e-03   7.2666269919203131e+01    1.6e+01    0.0e+00
   100     473    2.35e+00   7.2600584917043065e+01    3.4e+01    1.0e+00
   111     507    2.54e+00   7.2599790146012424e+01    6.5e+01    1.3e-04
warning: too many function evaluations (507)
        3    14001       10        5
 1: array(double,7507) min=0.783479 max=0.845731 avg=0.817507 std=0.01788
 ebv_star =      0.00900828
> 
\end{verbatim}
A typical output of {\tt sfit} is shown above, allowing the user to monitor the fitting process.
The values of the objective function are shown in the middle column, as the minimisation proceeds.
This is usuallty close to the $\chi^2$ of the fit. In this example, the fit is bad (the final $\chi^2 =72.6$), and the user would attempt to improve it. Guidelines for this are given in Sec. \ref{s:sense}.
Turning on the kinematics with the option {\tt kin=1} usually makes a huge difference.

\subsection{Options}

A number of options are available to enhance the fit, the most important one being the kinematics switch ({\tt kin=1}). All the options can be seen by typing:
\begin{verbatim}
> info,sfit
func sfit(gallist,base,&_ki,&_resi,nMC=,MC=,verb=,maxit=,maxitMC=,noskip=,
meval=,mevalMC=,meval3=,kin=,epar=,MASK=,sadguess=,guess=,mus=,vlim=,
losguesswidth=,sav=,asc=,RMASK=,zlim=,rmin=,bf=,padv=,frtol1=,N=,nor=,nde=,
sigdef=,L1=,L2=,L3=,L4=,muc=,co=,pl=,mucov=,cov=,parage=,RG=,muv=,mux=,muz=,
mue=,mub1=,mub2=,z2d=,bnpec=,splin=,__sigm=,AMRp=,muAMRp=,rrAMRp=,Nb=,lsp=,
plavgW=,nedgemask=,dMC=,MC_noise=,pr=,fg=,hidepaths=,piecewise_sad=)
\end{verbatim}

The most useful options are described in the following table. Others should be ignored and not played with.

\begin{tabular}{lp{9cm}}
{{\tt {kin=}}} & 1 to allow for non-parametric LOSVD determination. If 0 is used then the spectral models in the basis {\tt b} must already have the correct broadening and shift. \\
{\tt vlim=} & Velocity range on which the LOSVD is computed. An array of 2 elements must be provided, containing the lower and upper limits of the velocity range, such as {\tt vlim=[-1500.,1500.]}, in km/s. The values given as examples here are the default. \\
{{\tt noskip=}} &if 0 then spectra that already have a results file will not be processed. Useful when processing large datasets. 1 overrides this and reprocesses the spectrum even if it has already been processed before. \\
{\tt epar=} & type of extinction applied to the model. 1: extinction law parametered by E(B-V) taken from \cite{calzetti01}. 3: non-parametric extinction curve with {\tt nde} anchor points evenly distributed along the spectrum. This is usually a good choice because it absorbs smooth flux calibrations errors as well or errors in the correction of the instrument responds. default is 3. \\
{\tt mux=} & Smoothing parameter for the SAD. ($\mu_{\rm x}$ in the STECMAP paper). \\
{\tt muz=} & Smoothing parameter for the AMR. ($\mu_{\rm Z}$ in the STECMAP paper). \\
{\tt muv=} & Smoothing parameter for the LOSVD. ($\mu_{\rm v}$ in the STECKMAP paper). \\
{\tt mue=} & Smoothing parameter for the non-parametric extinction curve.  \\
{\tt nde=} & Number of anchor points for the non-parametric extinction curve. \\
{\tt mub1=} & 1st Binding parameter for the AMR ($\mu_C$ in the STECMAP paper). Do not modify. Wildly unstable behaviour may ensue. \\
{\tt mub2=} & 2nd Binding parameter for the AMR ($\mu_C$ in the STECMAP paper). Do not modify. Wildly unstable behaviour may ensue. \\
{\tt L1=} & Penalization for the SAD. Allowed values are {{\tt ``D1''}} (gradient), {\tt ''D2''} (Laplacian, as in STECMAP paper, or {\tt ``D3''}, 3rd order. Default is Laplacian. \\
{\tt L2=} & Penalisation for the LOSVD. Same comments as for {\tt L1}. \\
{\tt L3=} & Penalisation for the AMR. Same comments as above except default is gradient. \\
{\tt L4=} & Penalisation for the non-parametric extinction curve. Same comments as above, default is Laplacian.  \\
 \end{tabular}

\chapter{Reading the output files}
When the option sav$=1$ is given to sfit, the results are saved. The various results files are created in the same directory as the initial data file, upon success of the fitting procedure.
The .pdb file is the result of the convert\_all procedure. It is a binary file containing the initial data, a wavelength vector, an error spectrum and sometimes a mask if it has been supplied in the proper way (such as when dealing with SDSS data).
This chapter describes the other files available.
\section{Description of the results files}
The format of the files is text as follows:
\begin{enumerate}
\item{variable name}
\item{length of vector}
\item{values}
\item{variable name}
\item{length of vector}
\item{...}
\end{enumerate}

\subsection{Stellar content}
These are the files:  res-SAD, res-MASS, res-SFR and res-AMR.
They look like this (here for the res-MASS file):
\begin{verbatim}
 Ages (Myr)       10
        10.0000        23.2692        54.1455        125.992        293.173
        682.190        1587.40        3693.75        8595.06        20000.0
 Masses in each time bin       10
        39.4829        17.2968        5.35236        2.45517       0.997133
        5.72450        251.843        2432.88        16218.1        39665.3
\end{verbatim}

The variables relative to the stellar content all refer to a time axis, which can be seen as lookback time or the age of the stellar component. It is always given first in these files. The (Myr) indicates the time axis is in Myr.
Then comes the field of interest, here the masses. Here is a short description of the content of each file:
\begin{itemize}
\item{{\bf res-SAD}: The Stellar Age Distribution, i.e. the contribution in flux of each component to the observed spectrum. It is normalized so that the sum of the SAD is 1. It is the basic quantity STECKMAP is working with (rather than masses or SFR). In a typical STECKMAP run the SAD is shown in the top plot of the results panel.}
\item{{\bf res-AMR}: The Age-Metallicity Relation. This gives the metallicity $Z(i)$ of each component i of age $age(i)$, hence defining effectively an age-metallicity relation. In extragalactic astronomy, the metallicity is usually understood as a total abundance of metals, i.e. the fractional mass of everything which is not H or He. With this convention, in the models used here the metallicity of the sun is 0.02. Some authors will use 0.019 instead, but in practice this does not make a huge difference in the interpretation of the results, in particular as long as one does not embark into computing abundances of individual elements. In a typical STECKMAP run the AMR is shown as the middle plot of the results panel.}
\item{{\bf res-MASS}: This file gives the stellar mass as a function of the age.
Note that it is understood as an ``initial mass'', i.e. the mass the given component had at the time of its birth. No mass loss is taken into account here. The masses for each time bin $i$ are simply computed as:
\begin{equation}
mass(i)=\frac{sad(i)}{M/L(age(i),Z(i))}
\end{equation}
where $M/L(age,Z)$ is the mass to light ratio of a stellar population of age $age$ and metallicity $Z$. Again since no mass loss is taken into account this refers to the {\em initial mass}, i.e. it only accounts for the dimming of the population, not for the decrease/recycling of its stellar mass. The masses are given in principle in solar masses but the normalization is arbitrary: the relative masses (i.e. between the various bins) are correct but the actual normalization results from the fact that $sad$ is normalized to unity, i.e. $\sum_{i} sad(i)=1$.}
\item{{\bf res-SFR}: The Star Formation Rate (SFR) as a function of lookback time. It is obtained as:
\begin{equation}
SFR(i)=\frac{mass(i)}{\Delta t(i)} \, ,
\end{equation}
where $\Delta t(i)$ is the duration or extent of the age bin $i$. Note that this is {\em not} $t(i+1)-t(i)$ but is rather computed as like $\Delta t(i)=t(i+1/2)-t(i-1/2)$. The SFR is given in $M_{\odot}/yr$ but is not physically normalized. As for the masses, the relative variations are correct but not their absolute values, and the actual normalization is essentially inherited from that of the masses $mass$.} 
\end{itemize}

\subsection{Kinematics}
The broadening function of the data results from the convolution of the LOSVD and the instrument's PSF or LSF. However, I will often abusively call it the LOSVD, and talk about kinematics eventhough I should be talking about kinematics $+$ instrumental characteristics. The broadening function is stored in the {\bf res-LOSVD.txt} file.
It begins as follows:
\begin{verbatim}
 v (km/s)      369
       -1003.75       -998.323       -992.892       -987.460       -982.029
       -976.597       -971.165       -965.733       -960.301       -954.869
       ...
\end{verbatim}
This describes the velocity range (in km/s) on which the LOSVD has been computed. Next come the actual values of the broadening function for each of these velocities:
\begin{verbatim}
 g(v)      369
    1.57828e-06    1.62290e-06    1.65155e-06    1.65992e-06    1.63955e-06
    1.61256e-06    1.54141e-06    1.47136e-06    1.33613e-06    1.13671e-06
...
\end{verbatim}
The LOSVD or BF $g(v)$ is normalized to unity, i.e. its sum is equal to 1.

\subsection{Spectra}
A number of spectra are available in the {\bf res-spectra.txt} file. In order of appearance are the wavelength range (in Angstrom), the data, the best fitting model, the weight vector, and, if the option epar$=$3 has been used, the non-parametric extinction curve (maybe calling it non-parametrically adjusted continuum (i.e. NPAC would make more sense...). Although these are readily plotted in a usual STECKMAP run, this results file should allow the user more flexibility in displaying/examining his fit.

\chapter{Making sense of the results}
\label{s:sense}
This chapter is still work in progress. It gives a number of general recipes to improve the significance of the result, mostly by improving the fit and running Monte Carlo tests.
\section{How to improve your fit ?}
While a good $\chi^2$ (i.e. close to 1) does not necessarily mean that the derived SFH is the {\em true} SFH of the observed object, it is quite obvious that a SFH resulting from a bad fit, i.e. with a $\chi^2>1.2$ for instance (as a rule of thumb), can not be trusted. Hence, part of the process of using STECKMAP involves improving the fit, or lowering the $\chi^2$ of the fit. In some cases, the S/N of the data has been overestimated by the observer. It can be set manually using the {\tt SNR0=} keyword {\tt convert\_all} (see Sec. \ref{s:convert}). In such a case the user should also check afterwards that the residuals of the fit do indeed look like white noise.
The 2 other important things to investigate are the bad pixels or emission lines, and the hyperparameters.

\subsection{Masking bad pixels/emission lines}
Since the model libraries only contains stellar models with absorption lines, care must be taken in masking the nebular emission lines, possible sky subtraction and other bad pixels. A list of common emission lines in the optical is given in Appendix A.
A mask can be easily generated as a vector of pairs of wavelengths. The following mask is designed for masking a 10 \AA wide region around H$_\beta$ and H$_\alpha$ for a spectrum that in its restframe. The call to {\tt sfit} using the mask is also shown.
\begin{verbatim}
> mask=[[4856.,4866.],[6558.,6568.]]
> x=sfit(a,b,RMASK=mask)
\end{verbatim}

\subsection{Dependence on hyperparameters}
The default smoothing parameters usually achieve a good balance between keeping the solution smooth while still fitting the data well. From there, increasing {\tt mux} for instance will usually yield a smoother solution, at the price of a slightly larger $\chi^2$, as shown in \cite{stecmap,steckmap}.
On the other hand, lowering {\tt mux} will improve the $\chi^2$ is a negligible way, while making the SAD very unstable and sensitive to noise (as MonteCarlo experiments show). Lowering {\tt mux} further will at some point result in catastrophic failure of the minimisation procedure, and will return a totally chaotic SAD, with a bad $\chi^2$. 
It may be informative for the user to play around with the smoothing parameters (starting with {\tt mux}) in a range of a few decades ($10^{-2} - 10^2$) around the defaut values. It will also be useful to check out which values have been used in existing publications using STECKMAP (given in the foreword to this guide).

\subsection{Summary of guidelines for improving the fit}
In order of importance the following steps must be taken.
\begin{enumerate}
\item{Kinematics: if the kinematics are off ({\tt kin=0}) and the redshift and BF are incorrect, the fit can not be good. Turn the kinematics on ({\tt kin=1)} or correct manually the redshift and/or the BF (the models can be broadened to an arbitrary resolution using the keyword {\tt R=} in {\tt bRbasis3}.}
\item{Bad pixels/emission lines.}
\item{Re-estimation of the S/N of the data: this is usually a delicate task, although in some cases it can be obvious that the S/N is overestimated.}
\item{Investigate dependence on hyperparameters. Usually the default parameters are fine but in some very rare cases better $\chi^2$ can be obtained with non-default values.}
\end{enumerate}

\section{Assessing the significance of the results}
It is natural and wise to wonder about the sgnificance of our results. There are at least 3 types of errors we should care about:
\begin{itemize}
\item{Noise in the data: photon noise, read-out noise, background noise... }
\item{Methodological errors: altough authors of spectral inversion methods do not like to think about it there can be systematic effects due to the method itself and the choices that have been made in terms of interpolating between the ssp models, their normalization etc.... Perhaps the best way to investigate these is to use different methods in simple, controlled experiments, as was done in \cite{koleva08}.}
\item{Theoretical uncertainties: the ssp models are at the core of galaxy spectra inversion techniques, and they are not perfect (uncertainties in the stellar parameters and coverage of stellar libraries, uncertainties in stellar tracks...). A good way to investigate the effect of these is to always analyse a given spectrum with a collection of models rather than just one model. Several popular models are already plugged in the distributed version of STECKMAP (see table \ref{t:sspmodels}), and can be used simply by changing the parameter {\tt basisfile} in the {\tt bRbasis3} call. This approach was also used in \cite{koleva08}.}
\end{itemize}

In the following we will focus on the methods available in STECKMAP to tackle the first item and some of the second, via MonteCalro experiments.

\subsection{Robustness of the results: running Monte Carlo experiments}
Here we understand Monte Carlo experiments as: rerunning the inversion on the dataset with a randomized element. The randomized element in STECKMAP can either be the noise in the data or the first guess (i.e.e the starting point of the minimization method).

\subsection{Robustness wrt noise in the data}
Here we want to invert mock data, which will be either the original data with a different realization of the noise, or the best model we have of that data, with noise added according to the noise vector. The relevant options for {\tt sfit} are listed in Tab. \ref{t:optionsnoiseMC}

\begin{table}[h]
\begin{tabular}{|l|p{11cm}|}
\hline
{\tt nMC=} & number of MC experiments to be performed \\ 
\hline
{\tt dMC=} & determines the mock data to be used for the MC experiments. Two choices are possible only:
 \\
& {\tt dMC=''bestmodel''} will create a mock spectrum by taking the best model (corresponding to the derived solution SAD, AMR, BF etc...) and noising it using the noise spectrum. The latter has been provided by the user during the conversion with {\tt convert\_all}. See Sec. \ref{s:convert} for the specifications of the noise spectrum or global signal to noise ratio. \\
& {\tt dMC=''data''}. In this case the mock spectrum will be  the original data with noise added according to the noise spectrum or the global SNR. Note that with {\tt dMC=''bestmodel''} we expect $\chi^2 \approx 1$ for the MC runs, while in the case {\tt dMC=''data''}, we expect $\chi^2 \approx 2$ because the original data already contains noise.\\
\hline
{\tt MC\_noise=} & {\tt ``yes''}: mock data is noised according to noise spectrum or global SNR\\
{\tt MC\_noise=} & {\tt ``no''}:  mock data is not noised (if mock data is the original data then no extra noise is added, same for best model)\\
\hline
\end{tabular}
\caption{Options for {\tt sfit} for performing MonteCarlo experiments to test the robustness of the results with respect to noise in the data.}
\label{t:optionsnoiseMC}
\end{table}

A typical call will look like:\\
{\tt x=sfit(a,b,kin=1,noskip=1,sav=1,nMC=5,dMC=''bestmodel'',MC\_noise=''yes'')}

\subsection{Robustness wrt first guess choice}
All iterative optimization methods require a starting point, which is usually referred to as the {\em first guess}. In the case of SFH reconstruction from spectra, scientists aim at designing a method and an implemetation of it, that will be as little as possible sensitive to the first guess. Ideally, we would like STECKMAP to give always the same answer, no matter what the first guess was. In STECKMAP, the default first guess is a flat distribution for all fields. This flat first guess has 2 good properties.
\begin{itemize}
\item{Convergence: the method usually converges to a good solution (in the $\chi^2$ and regularization sense), which is NOT TRUE for all first guesses: in fact, it is fairly easy to come up with a first guess that leads to a secondary minimum with a very bad $\chi^2$, or which leads to divergence of the minimization method.}
\item{Uniformity}
\end{itemize}

From my own experience using STECKMAP I acquired confidence in the fact that the flat first guess {\em usually} does a good job. However there is no proof that it is always the best or that it will not sometimes lead to misfits or secondary minima. Note that this is true for all available methods \cite{ulyss,moped01}. To cope with this, many authors note the fact that in order to make sure that they reach a ``global'' minimum and not a secondary minimum, they have to run the fitting engine several times, using a family of first guesses.

Several techniques to vary the first guess have been implemented in {\tt sfit}. They are activated and controlled using the options of Tab. \ref{t:optionsfg}. They complete the options given in Tab. \ref{t:optionsnoiseMC}, which affect the mock data. The preferred setup for first guess variation experiments should be : {\tt dMC=''data'',MC\_noise=''no'')}.

\begin{table}[h]
\begin{tabular}{|l|p{11cm}|}
\hline
{\tt nMC=} & number of MC experiments to be performed \\ 
\hline
{\tt fg=} & determines the type of first guess to be used for the MC experiments. Three choices are possible:
 \\
{\tt fg=''flat''}& will create a flat first guess. This is the default. \\
{\tt fg=''RB''}& stands for ``random burst''. The first guess for the SAD is a single burst at a random age within the ages of the SSP basis. The width of the burst is 3 age bins. The AMR is flat as well with a random metallicity within the Z range of the SSP basis. The color excess is 0 or the NPEC vector is 1 everywhere, depending on the option chosen to deal with extinction, and the BF is a gaussian centered on 0 km/s.\\
{\tt fg=''RG''}& stands for ``random guess''. All fields are randomized. Not very stable. Prefer the 2 first alternatives.\\
\hline
\end{tabular}
\caption{Options of {\tt sfit} for performing MonteCarlo experiments to test the robustness of the results with respect to variations of the first guess.}
\label{t:optionsfg}
\end{table}

A typical call in order to test 5 random bursts as first guesses, fitting the original data (and not some mock data), will look like:\\
{\tt x=sfit(a,b,kin=1,noskip=1,sav=1,nMC=5,dMC=''data'',fg=''RB'',MC\_noise=''no'')}

\subsection{Output of MonteCarlo experiments}
The results of each experiments are conveniently stored in the {\tt -res.txt} files, using the same format as the non-MC results, so that for a {\tt nMC=5} run, one has 6 SADs, AMRs, etc...
At the moment the corresponding first guesses are not stored in a txt file (altho thats the way it should be!). Here are a few lines of example showing how to recover them. In this example all the fields are concatenated and not properly normalized. It is nonetheless enough to see where the first guess SADs peaks.
\begin{verbatim}
#include "STECKMAP/Pierre/POP/sfit.i"
a=convert_all(fV,z0=0.0021);
b=bRbasis3(,basisfile="MILES");
x=sfit(a,b,noskip=1,kin=1,nMC=5,dMC="data",fg="RB",MC_noise="no",sav=1,meval=500);
// plot the first guesses (all fields concatenated)
upload,a.resfile(1);
ws,1;
plb,RBs;
\end{verbatim}

\chapter{Tweaking STECKMAP}
In some cases you will want to use STECKMAP in a way that was not initially planned by the author. This chapter gives a few tricks and corresponding yorick codes to do so.
\section{Adding a constraint on the AMR}
\section{Fitting the spectrum with SSPs $+$ power-law}






\bibliographystyle{plain}
\bibliography{mybib.bib}
%

\appendix

\chapter{Common emission lines in the optical}
Emission lines of a variety of chemical elements are usually seen in optical spectra of galaxies. Here is a compilation of the most common ones. It is not meant to be exhaustive but will help the user to design masks efficiently. Wavelengths are given in the air.

\begin{table}[h]
\begin{center}
\begin{tabular}{ccc}
\hline
Emission  & Wavelength & Ref.\\
line & (air, \AA) & \\
\hline
\hline
H$\alpha$& 6563 & (2)\\
H$\beta$& 4861 & (2)\\
H$\gamma$& 4340 & (2) \\
H$\delta$& 4101 & (2)\\
H$\epsilon$& 3970 & (2)\\
H$\zeta$& 3889 & (2)\\
HeI & 5876 &	(1)\\
$[$O I$]$& 6300 &	(1)\\
$[$O II$]$& 3727  & (1) \\
$[$O III$]$& 4363 &	(1)\\
$[$O III$]$& 4959 &	(1)\\
$[$O III$]$& 5007 &	(1)\\
$[$N II$]$& 6584 &	(1)\\
$[$N II$]$& 6548 &	(1)\\
$[$Ne III$]$& 3869& (1)\\
$[$S II$]$& 6716 &	(1)\\
$[$S II$]$& 6731 &   (1)\\
\hline
\end{tabular}
\caption{Common emission lines in optical spectra.
References: 
(1): http://www.mpa-garching.mpg.de/SDSS/DR7/raw\_data.html, (2): http://en.wikipedia.org/wiki/Balmer\_series
}
\end{center}
\end{table}

\label{lastpage}
\end{document}